\documentclass[12pt,preprint]{aastex}

\begin{document}

\title{Optical flares and flaring oscillations on
the M-type eclipsing binary CU Cnc}

\author{Qian S.-B.\altaffilmark{1,2,3}, Zhang J.\altaffilmark{1,2,3}, Zhu
L.-Y.\altaffilmark{1,2}, Liu L.\altaffilmark{1,2}, Liao W.-P.\altaffilmark{1,2}, Zhao E.-G.\altaffilmark{1,2}, He J.-J.\altaffilmark{1,2}, Li L.-J.\altaffilmark{1,2,3}, Li
K.\altaffilmark{1,2,3} and Dai Z.-B.\altaffilmark{1,2}}

\altaffiltext{1}{National Astronomical Observatories/Yunnan Observatory, Chinese Academy of Sciences, P.O. Box 110, 650011 Kunming, P. R. China (qsb@ynao.ac.cn)}

\altaffiltext{2}{Key laboratory of the structure and evolution of celestial objects, Chinese Academy of Sciences, P.O. Box 110, 650011 Kunming, P. R. China}

\altaffiltext{3}{Graduate University of the Chinese Academy of Sciences, 100049 Beijing, P. R. China Yuquan Road 19\#, Sijingshang Block, 100049 Beijing City, P. R. China}

\begin{abstract}
We report here the discovery of an optical flare observed in R band
from the red-dwarf eclipsing binary CU Cnc whose component stars are
at the upper boundary of full convection ($M_1=0.43$ and
$M_2=0.4$\,$M_{\odot}$, $M_{\odot}$ is the solar mass). The
amplitude of the flare is the largest among those detected in R band
($\sim{0.52}$\,mag) and the duration time is about 73 minutes. As
those observed on the Sun, quasi-periodic oscillations were seen
during and after the flare. Three more R-band flares were found by
follow up monitoring. In total, this binary was monitored
photometrically by using R filter for 79.9\,hours, which reveals a
R-band flare rate about 0.05 flares per hour. These detections
together with other strong chromospheric and coronal activities,
i.e., very strong $H_\alpha$ and $H_\beta$ emission features and an
EUV and X-ray source, indicate that it has very strong magnetic
activity. Therefore, the apparent faintness ($\sim$1.4 magnitude in
V) of CU Cnc compared with other single red dwarfs of the same mass
can be plausibly explained by the high coverage of the dark spots.
\end{abstract}

\keywords{Stars: binaries : close --
          Stars: binaries : eclipsing --
          Stars: individuals (CU Cnc) --
          Stars: activity --
          Stars: flare --
          Stars: late-type}

\section{Introduction}

Though M-type stars (red dwarfs with mass less than
$0.6$\,$M_{\odot}$, and $M_{\odot}$ is the solar mass) are the most
populous stellar objects in the Galaxy, to date, only about a dozen
of eclipsing red-dwarf binaries in detached systems was detected and
studied in details because of the low probability of finding them
and their very low intrinsic brightness (e.g., Blake et al. 2008;
Irwin et al. 2009; Torres \& Ribas 2002; Morales et al. 2009).
Fundamental stellar properties of these binaries, especially masses
and radii, can be measured in a high precision (better than 2\%).
Therefore, they play a key role in our understanding of stellar
physics near the bottom of main sequence in the Hertzsprung-Russell
diagram.

With an orbital period of 2.771468 days, CU Cancri (=GJ\,2069A,
HIP\,41824) is one of a few extremely low-mass eclipsing binaries
(e.g., Ribas, 2003; Blake et al., 2008; Irwin et al., 2009; Morales
et al., 2009). It is a nearby (d=12.8\,PC) 11.9-mag spectroscopic
binary system with M3.5\,Ve components, where the "e" refers to the
Balmer lines being in emission in the "quiescent" state (e.g., Reid
et al., 1995; Delfosse et al., 1999b). Photometric monitoring with
the 0.7-m Swiss telescope at the European Southern Observatory
revealed that it is an eclipsing binary (Delfosse et al. 1999a).
Subsequent investigations suggested that the eclipsing binary has a
fainter visual companion at an angular distance of $\sim12"$, with
common proper motion (Giclas et al. 1959) and radial velocity
(Delfosse et al. 1999b). This fainter companion is also a double
system.

Complete light curves in R and I bands were published by Ribas
(2003) who determined absolute parameters of CU Cnc by combining the
photometric elements with the existing spectroscopic solutions. The
derived orbital inclination was $i=86.^{\circ}34(\pm0.03)$
indicating that it is a total eclipsing binary and absolute
parameters can be determined in high precisions. Their results were:
$M_1=0.4333(\pm0.0017)$\,$M_{\odot}$,
$M_2=0.3980(\pm0.0014)$\,$M_{\odot}$,
$R_1=0.4317(\pm0.0052)$\,$R_{\odot}$, and
$R_2=0.3908(\pm0.0094)$\,$R_{\odot}$. It was found that the two
components in the binary are fainter than other stars of the same
mass with a magnitude difference about 1.4\,mag in V and 0.35 mag in
K band. As discovered in other eclipsing binary stars, most
theoretically evolutionary models underestimate the radii of the
component stars by as much as 10\%.

High resolution spectra were obtained by Ribas (2003) that revealed
very strong $H_{\alpha}$ and $H_{\beta}$ emission features (double
lines). As that of YY Gem, the $H_{\alpha}$ equivalent width of CU
Cnc is rather large but not unreasonable when compared with young
red dwarfs (e.g., Soderblom et al. 1991). However, the strong
$H_{\alpha}$ emission in the two binary systems is more related to
the tidally spin-up caused by orbital synchronization rather than
age (e.g., Ribas 2003). Apart from showing $H_{\alpha}$
chromospheric emission, CU Cnc also displays strong of EUV and X-ray
emission (e.g., Voges et al. 1999; Schmitt et al. 1995). The
calculation of $L_{X}/L_{Bol}$ by Ribas (2003) yielded a value of
$10^{-3}$ that is very close to the value of YY Gem. As for the
dark-spot activity of CU Cnc, the analysis of the light curve by
Ribas (2003) indicated that there are two spots -- one on each
component. The spot on the primary component appears to be
relatively small (with a radius of $9^{\circ}$ and 450 K cooler than
the photosphere), while the secondary component has a much larger
spot (with a radius of $31^{\circ}$ and a temperature difference
with the surrounding photosphere of 200 K).

Flares are known as sudden and violent events that release magnetic
energy and hot plasma from the stellar atmospheres. They are
observed on magnetically active stars and, much more closely, on the
Sun. Observations have shown that flares in M-type stars occur more
frequently than those on G- and K-type stars (e.g., Moffett 1974;
Lacy et al. 1976; Henry \& Newsom 1996). However, no flare
activities on CU Cnc were observed. Here, we report the flares of CU
Cnc observed in R band including a flare with the largest amplitude
and showing quasi-periodic oscillations.

\section{Optical flares from CU Cnc}

To understand the properties of the variations of the light curves
and the orbital period changes of red-dwarf eclipsing binaries
(e.g., YY Gem, TU Boo, CU Cnc, FS Aur, NSVS\,02502726, and DV Psc),
we were monitoring them photometrically by using two Andor DW436\,2K
CCD cameras amounted on the 1.0-m and 60-cm telescopes in Yunnan
observatory. During the monitoring of CU Cnc with the 60-cm
telescope on October 28, 2009, we were lucky to detect a large flare
in the R-band (see in Fig. 1). The integration time for each CCD
image was 90\,s. Coordinates of the comparison and the check stars
were $\alpha_{2000}=08^{h}31^{m}37.4^{s}$ and
$\delta_{2000}=+19^\circ23'49.5"$ for the comparison, and
$\alpha_{2000}=08^{h}32^{m}09.7^{s}$ and
$\delta_{2000}=+19^\circ26'59.3"$ for the check star. The observed
images were reduced by using PHOT (measure magnitudes for a list of
stars) of the aperture photometry package of IRAF.

As shown in Fig. 1, the amplitude of the flare is about 0.52\,mag
and the duration time is about 73 minutes. Its characteristic shape
shows a rapid brightness increase (the impulsive phase) followed by
a gradual decline, which are similar to those seen in solar flares.
This flare was occurring during the primary eclipse (from phase
0.9935 to phase 1) indicating that it is most probably from the
secondary component. Meantime, due to high time resolution of our
data, three small eruptions were observed to be superimposed on the
flare, which are more clearly seen in the lower panel of Fig. 1. The
duration time of each eruption is about 3 minutes and the mean
amplitude is about 0.046 magnitudes. These properties indicate that
there are quasi-period pulsations in the flare, which were often
observed in solar flares with periods ranging from fraction of
seconds to several minutes (e.g., Nakariakov \& Melnikov 2009). The
mean amplitude of the quasi-period pulsations is about 4.6 times
larger than the photometric error (about $0.^{m}01$) suggesting that
it is not from random noise in the light curve. After this flare,
four smaller eruptions were observed subsequently indicating that
there are post-flare quasi-period pulsations, which were also
observed in subsequent solar flares (e.g., Grechnev et al., 2003).
The maximum relative luminosity at the flare peak can be calculated
by using,
\begin{equation}
L_{Max}=2.5^{-\Delta{m_A}}\times{L_{0}},
\end{equation}
where $L_{0}$ is the quiescent luminosity of the binary and
$\Delta{m_A}$ is the amplitude of the flare. It is shown that at the
peak of the flare the luminosity of the binary increases by 1.61
times. By considering that both component stars in CU Cnc are
similar, the luminosity of the flaring component should increase by
about 3.22 times.

\begin{figure}
\begin{center}
\includegraphics[angle=0,scale=.8 ]{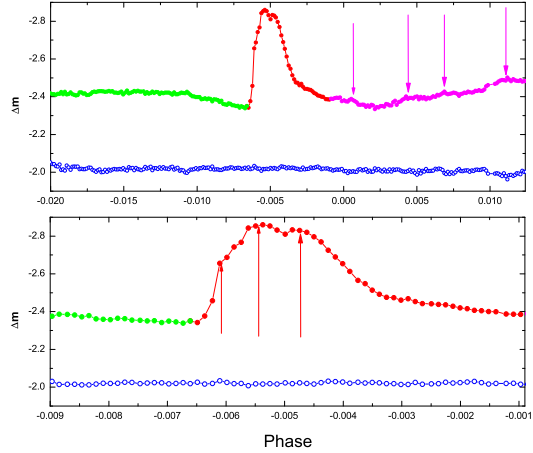}\caption{A very
energetic R-band flare from CU Cnc. All of the solid dots in
different colours represent the magnitude difference between CU
Cancri and the comparison star. The energetic flare (red dots) was
observed on October 28, 2009 and occurred during the primary eclipse
indicating that it may be from the secondary component. Green solid
dots refer to the brightness in "quiescent" state, while blue open
circles represent the difference between the comparison and the
check stars. Magenta arrows in the upper panel show the times of the
four smaller post-flare eruptions. Details of the properties of the
flare are displayed in the lower panel where three small eruptions
(the arrows) are detected during the flare suggesting there is a
quasi-periodic oscillation in the flare. Phases of the observation
were calculated with the linear ephemeris given by Ribas (2003):
$T(Min. I)=HJD\,2450208.5068 +2.771468\times{E}$.}
\end{center}
\end{figure}

Three more flares in R band were found by following up monitoring
with the 60-cm telescope. We used the following criteria. (a) A
flare should last for several minutes and contain more than one data
point, since a peak represented by a single data point could be due
to cosmic rays from the direction of the stars, and (b) the
amplitude of the peak should be no less than $0.03^{m}$ (three times
of the photometric error about $0.^{m}01$). The one observed on
December 10, 2009 occurred just after the secondary eclipse (as
shown in Fig. 2), and the other two took place out of eclipses. We
do not know the three R-band flares are from the more massive
component or the less massive one. All of the three flares are shown
in Fig. 2 where phases of the observation were computed based on the
ephemeris given by Ribas (2003).

In all, we were monitoring CU Cnc for 90.8\,hours. Meanwhile, R
filter was used for 79.9 hours revealing a R-band flare rate of 0.05
flares per hour. The log of the photometric monitor of CU Cnc is
listed in Table 1. The flare rate of CU Cnc is comparable with that
of the other cool eclipsing binary star CM Dra (e.g., Kim et al.
1997). However, since the four flares were observed in R band
indicating that they are energetic ones. We suspect that the true
flare rate might be higher than this value because some less
energetic flares can not be found in R band.

\begin{table*}
\caption{The log of the photometric monitor for CU Cnc.}
\begin{center}
\begin{tabular}{llllll}\hline\hline
Date & Start UT & End UT & Filter & Flare & Telescopes\\\hline
8 December 2007 & 19:00:31 & 19:58:03 & R  & No  & The 1.0-m\\
28 March 2008   & 14:56:31 & 17:39:51 & I  & No  & The 1.0-m\\
28 October 2009 & 20:04:29 & 22:34:18 & R  & Yes & The 60-cm \\
22 November 2009& 18:35:59 & 20:29:45 & R  & No  & The 60-cm \\
28 November 2009& 16:57:26 & 22:14:32 & V  & No  & The 1.0-m\\
29 November 2009& 16:36:07 & 19:18:42 & R  & No  & The 60-cm \\
30 November 2009& 17:13:44 & 18:15:13 & R  & No  & The 60-cm \\
1 December 2009 & 16:24:45 & 22:31:08 & R  & No  & The 60-cm \\
3 December 2009 & 16:50:52 & 22:02:32 & V  & No  & The 1.0-m\\
7 December 2009 & 16:13:10 & 20:29:47 & R  & No  & The 60-cm \\
8 December 2009 & 15:58:09 & 22:50:54 & R  & Yes & The 60-cm \\
9 December 2009 & 16:10:47 & 22:59:43 & R  & No  & The 60-cm \\
10 December 2009& 16:08:36 & 22:53:47 & R  & Yes & The 60-cm \\
11 December 2009& 17:09:30 & 22:48:37 & R  & No  & The 60-cm \\
12 December 2009& 15:46:17 & 23:03:46 & R  & No  & The 60-cm \\
13 December 2009& 16:16:59 & 20:18:19 & R  & Yes & The 60-cm \\
14 December 2009& 15:40:24 & 21:53:20 & R  & No  & The 60-cm \\
15 December 2009& 15:38:29 & 23:02:50 & R  & No  & The 60-cm \\
16 December 2009& 15:38:04 & 20:09:29 & R  & No  & The 60-cm \\
17 December 2009& 15:38:09 & 20:40:32 & R  & No  & The 60-cm \\
\hline
\end{tabular}
\end{center}
\end{table*}

\begin{figure}
\begin{center}
\includegraphics[angle=0,scale=.8 ]{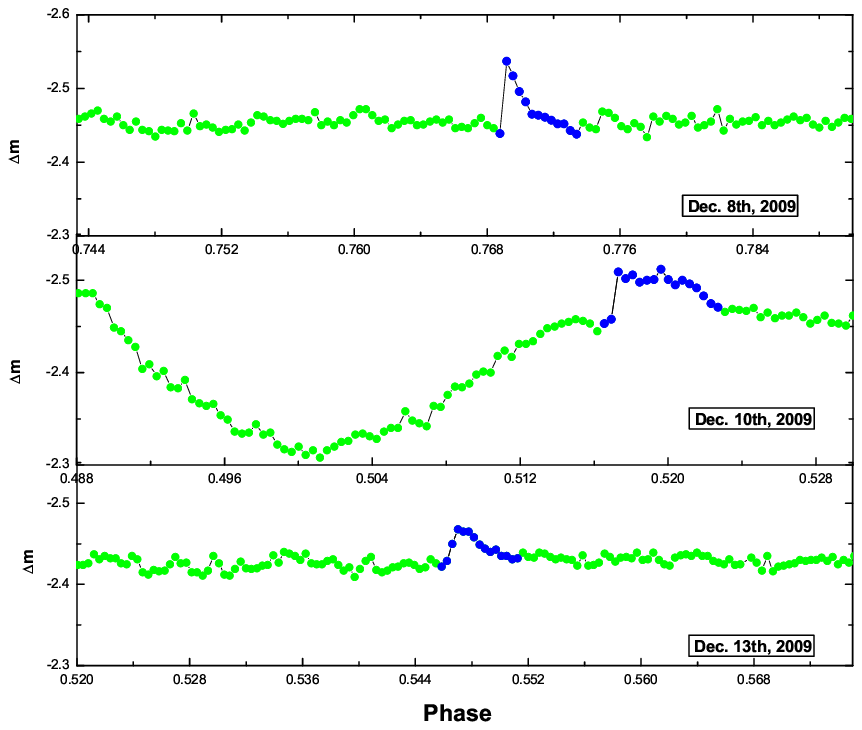}\caption{Three
R-band flares of CU Cnc observed on December 8, 10, and 13, 2009.
Phases of the observation were calculated by using the linear
ephemeris given by Ribas (2003). }\end{center}
\end{figure}

\section{Discussions and conclusions}

Since flares are most readily observed in U and B bands, the fact
that the four flares were observed in R-band suggests that they are
highly energetic. Among all of the R-band flares observed so far
(see Table 2), the one from CU Cnc that took place on October 28,
2009 has the largest amplitude indicating that it is one of the most
energetic giant flares. The detection of four R-band flares
especially the very energetic one from CU Cnc reveals that it is a
strong magnetically active system.

\begin{table*}
\caption{Flares of cool stars observed in R-band.}
\begin{center}
\begin{tabular}{llll}\hline\hline
Star names & Flare amplitude & Total duration & Ref.\\
           & $\Delta{m}$     & Minutes        &  \\\hline
V405 And &  0.12  & 80 & Vida et al. (2009)\\
CU Cnc   &  0.52  & 73 & The present authors\\
CU Cnc   &  0.10  & 17 & The present authors\\
CU Cnc   &  0.05  & 23 & The present authors\\
CU Cnc   &  0.04  & 38 & The present authors\\
FR Cnc   &  0.21  & 41 & Golovin et al. (2007)\\
WY Cnc   &  0.045 & 64 & Kozhevnikova et al. (2006)\\
CM Dra   &  0.23  & 60 & Nelson \& Caton (2007)\\
CM Dra   &  0.04  &135 & Nelson \& Caton (2007)\\
CM Dra   &  0.08  &135 & Nelson \& Caton (2007)\\
CM Dra   &  0.09  &135 & Nelson \& Caton (2007)\\
CM Dra   &  0.02  & 34 & Nelson \& Caton (2007)\\
CM Dra   &  0.02  & 40 & Nelson \& Caton (2007)\\
CM Dra   &  0.21  & 92 & Kozhevnikova et al. (2004)\\
CM Dra   &  0.10  & 70 & Kozhevnikova et al. (2004)\\
CM Dra   &  0.03  & 21 & Kozhevnikova et al. (2004)\\
CM Dra   &  0.21  & 34 & Kozhevnikova et al. (2004)\\
DV PSc   &  0.02  &13.5& Zhang et al. (2010)\\
XY UMa   &  0.04  & 30 & Zeilik et al. (1982)\\
 \hline
\end{tabular}
\end{center}
\end{table*}

Quasi-periodic pulsations are a common feature of the flare energy
release process of the Sun that can be seen in all observational
bands with periods from a fraction of a second to several minutes
(e.g., Nakariakov \& Melnikov 2009). As for the stellar flares, a
few cases of quasi-periodic oscillations were reported. Rodon\`{o}
(1974) detected white light intensity oscillations (13\,s) during a
flare on the red-dwarf star II Tau. Mathioudakis et al. (2003)
observed 5 minutes white light intensity oscillations in a flare on
the RS CVn-type eclipsing binary II Peg. During of the observation,
quasi-period pulsations with a mean period of 3 minutes were
discovered on the largest-amplitude flare. Observational evidence
revealed that there exist post-flare oscillations. It is the first
time to detect quasi-periodic oscillations during and after the
flaring energy release of an close eclipsing binary star, and the
physical mechanisms to cause it are unclear.

At least 80\% of all stars in the Galaxy are Red dwarfs. However,
physical properties of these most common stars are poorly
understood. The serious problem in this field is the significant
discrepancy between the theoretical and observational mass-radius
relations, i.e., the observed radius is about 10\% larger than that
computed from theoretical models (e.g., Blake et al. 2008; Irwin et
al. 2009; Torres \& Ribas 2002; Morales et al. 2009). Some authors
have realized that this discrepancy could attribute to the strong
magnetic activities of the components in short-period red-dwarf
eclipsing binaries (e.g., Mullan \& Macdonald 2001; Chabrier et al.
2007; L\'{o}pez-Morales 2007; Torres et al. 2010; Morales et al.
2008, 2010; Devor et al. 2008). Mullan \& Macdonald (2001)
investigated the effects of magnetic field on stellar structure and
pointed out that their magnetic models predicted active M-type
dwarfs tend to have larger radii.

To resolve the mass-radius discrepancy, two scenarios were
considered (e.g., Chabrier et al. 2007). One is that magnetic field
and rotation can induce reduction of the efficiency of large-scale
thermal convection in the interior and thus can lead to less
efficient heat transport. The other is that the magnetic dark-spot
coverage decreases the star's radiating surface and also yields a
smaller effective temperature and a larger radius. Either one or a
combination of the two can predict larger radius than standard
stellar models, but the effect of dark spots is significant over the
entire low-mass domain (e.g., Chabrier et al. 2007; Morales et al.
2010). However, it is shown that high spot coverage (up to 50-100\%)
is needed to be assumed to solve the significant discrepancy
(Morales et al. 2010). On the observational aspect, to explain the
observations of M-dwarfs in the young open cluster NGC\,2516, the
required spot coverage is about 50 per cent in rapidly rotating
M-type dwarfs (e.g., Jackson et al., 2009). Reiners et al. (2009)
claimed very high filling factors of magnetic field in rapidly
rotating M-dwarfs.

In the mass-absolute magnitude diagram, CU Cnc was found to be
fainter than other stars of the same mass with a magnitude
difference about 1.4 mag in V band. The apparent faintness was
explained by Ribas (2003) as: (i) its components are some 10\%
cooler than similar-mass stars or (ii) there is some source of
circumstellar dust absorption. The detection of the four R-band
giant flares together with other strong chromospheric and coronal
emissions (see in Section 1) support the conclusion that the
faintness of the stars in CU Cnc is caused by groups of dark spots
via the enhanced magnetic activity of the cool components. The total
luminosity of a spotted star is given as (e.g., Chabrier et al.
2007; Morales et al. 2010),
\begin{equation}
L=S[(1-\beta)\sigma{T^4}+\beta\sigma{T_{s}^4}],
\end{equation}
where $\sigma$ is Stefan-Boltzmann constant, $T_s$ the temperature
of the spotted area, $T$ the temperature of the immaculate surface,
and $\beta=\frac{S_s}{S}$ is the fraction of stellar surface covered
by dark spots, where ${S_s}$ is the spotted area and ${S}$ is total
area of the active star. As pointed out by Ribas (2003), no
empirical information is available on the temperature of the spots.
Photometric solutions of CU Cnc derived by Ribas (2003) indicate
that the spot on the primary of CU Cnc is 450\,K cooler than the
photosphere, while the one on the secondary has a temperature
difference with the surrounding photosphere of $\sim$\,200\,K.
During the modulating of the light curves of M-type eclipsing
binaries, Morales et al. (2010) assumed the temperature of spots is
about $\sim$\,500\,K cooler than the photosphere. Here, by assuming
$\frac{T_s}{T}=0.85$, the faintness of CU Cnc can be explained by
93\% of dark-spot coverage on the photospheric surfaces (by
considering the 'mean' component). If we take into account that the
radius changes as the star becomes spotted, the coverage of the dark
spots should be larger. Observations of CU Cnc may provide
observational evidence for highly dark-spot coverage on component
stars in short-period M-type eclipsing binary stars, and thus the
mass-radius problem can be resolved.

However, by modeling the light curve of CU Cnc, Ribas (2003)
obtained a significantly lower spot coverage than our results. This
may be caused by the fact that the coverage of dark spots for
red-dwarf eclipsing binaries may be underestimated during the
photometric solution because photometric changes of eclipsing
binaries applied to determine the spot parameters are only sensitive
to the contrast between areas with different effective temperatures
and not to the total surface covered by spots (Morales et al. 2010).
On the other hand, perhaps axial symmetry of the spots also plays a
role. Both components in CU Cnc are at the upper boundary of full
convection. M-type dwarfs near the boundary like the two in CU Cnc
exhibit large-scale magnetic fields which are very strong
axisymmetric poloidal and nearly dipolar with very little temporal
variations (Morin et al. 2010). It is possible that the stable and
enhanced large-scale magnetic fields associate with the extremely
high dark-spot coverage discovered in CU Cnc.

As shown in Fig. 2, the out-of-eclipse brightness of CU Cnc is
changed by 0.035\,magnitude from Dec. 8 to 13, 2009. This is
evidence for dark-spot activity revealing a change of dark-spot
coverage about 6.7\%. Because of the tidal locking and the effect of
the close companions, the magnetic activity levels of close binaries
are much higher than those of single stars with the same mass.
Strong magnetic activity may significantly affect the structure and
evolution of eclipsing binaries and the physical properties of these
objects depart from the single ones (e.g., Mullan \& Macdonald 2001;
Chabrier et al. 2007).

\begin{figure}
\begin{center}
\includegraphics[angle=0,scale=.8 ]{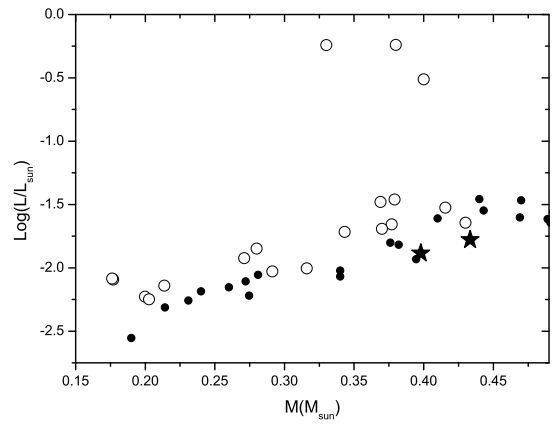}
\caption{Mass-luminosity relation for M-type stars with similar to
the components in CU Cnc. Solid dots represent to the eclipsing
binaries, while open circles refer to single stars. The positions of
the components in CU Cnc are shown as solid stars. }\end{center}
\end{figure}

To check whether similar faintness were observed in other M-type
eclipsing binary stars or not, we have compared parameters of some
eclipsing binaries (Kraus et al. 2011; Irwin et al. 2009, 2011;
Blake et al. 2008; Morales et al. 2009; Vaccaro et al. 2007;
L\'{o}pez-Morales \& Ribas 2005; L\'{o}pez-Morales et al. 2006;
Torres et al. 2002; Ribas et al. 2003; Hebb et al. 2006; Dimitrov \&
Kjurkchieva 2010; Moceroni et al. 2004; Creevey et al. 2005; Devor
et al. 2008). The mass-luminosity relation is shown in Fig. 3. For
comparison, parameters of some single stars given by Xia et al.
(2008) and Preibisch \& Mamajek (2008) are also shown in the figure
where solid dots refer to the eclipsing binaries, while open circles
represent single stars. Also display in this figure as solid stars
are the positions of the components in CU Cnc. The masses of those
single stars were estimated with their colors (or spectral types)
and magnitudes by considering their evolutionary tracks (e.g.,
Baraffe et al. 1998). As displayed in Fig. 3, even those single
stars are still young and magnetically active, they are usually
brighter than the components of short-period eclipsing binaries with
the same mass indicating highly dark-spot coverage on the component
stars. However, since the masses have been measured/estimated in
completely different ways, the final conclusion that eclipsing
binaries are fainter at the same mass would be premature. Moreover,
the photometric solutions of the eclipsing binaries were obtained
with several different software packages (e.g., Wilson \& Devinney
1971; Popper \& Etzel 1981; Southworth et al. 2004; Prsa \& Zwitter
2005). Further brightness comparison between M-type components in
eclipsing binaries and single red dwarfs and uniform solutions of
eclipsing binaries with the same method are required in the future.

\acknowledgments{
This work is partly supported by Chinese Natural Science Foundation
(No.11133007, No.10973037, No.10903026, and No.11003040) and by West Light
Foundation of the Chinese Academy of Sciences. New CCD photometric
observations of the system were obtained with the 60-cm and the
1.0-m telescopes in Yunnan Astronomical Observatory.}

\end{document}